\documentclass[11pt]{article}

\usepackage{acl}

\usepackage{times}
\usepackage{latexsym}

\usepackage[T1]{fontenc}

\usepackage[utf8]{inputenc}

\usepackage{microtype}

\usepackage{inconsolata}

\usepackage{graphicx}
\usepackage{amsmath}
\usepackage{subcaption}
\usepackage{todonotes}
\usepackage{tabularx}  
\usepackage{booktabs}  
\usepackage{amssymb} 
\usepackage{multirow}

\usepackage{algorithm}
\usepackage{algpseudocode}  
\usepackage[table]{xcolor}
\usepackage{tcolorbox}
\tcbuselibrary{listings,skins,breakable} 
\usepackage{listings}

\newtcblisting{promptbox}[1][]{
  colback=gray!5!white,
  colframe=gray!75!black,
  breakable,
  listing only,            
  listing engine=listings, 
  title=#1,
  listing options={
    basicstyle=\ttfamily\small,
    columns=fullflexible,
    keepspaces=true,
    showstringspaces=false,
    breaklines=true,
    numbers=none
  }
}

\title{Compact Multimodal Language Models as Robust OCR Alternatives for Noisy Textual Clinical Reports}

\author{
  \textbf{Nikita Neveditsin\textsuperscript{1}},
  \textbf{Pawan Lingras\textsuperscript{1}},
  \textbf{Salil Patil, M.Ch.\textsuperscript{2}},
  \\ 
  \textbf{Swarup Patil, M.D.\textsuperscript{2}},
  \textbf{Vijay Mago\textsuperscript{3}}
  \\
  \\ 
  \textsuperscript{1}Saint Mary's University, Halifax, Canada,
  \textsuperscript{2}Dhanwantari Hospital, Pune, India,
  \\ 
  \textsuperscript{3}York University, Toronto, Canada
  \\
  \small{
    \textbf{Correspondence:} \href{mailto:nikita.neveditsin@smu.ca}{nikita.neveditsin@smu.ca}
  }
}

\begin{document}
\maketitle

\begin{abstract}
Digitization of medical records often relies on smartphone photographs of printed reports, producing images degraded by blur, shadows, and other noise. Conventional OCR systems, optimized for clean scans, perform poorly under such real-world conditions. This study evaluates compact multimodal language models as privacy-preserving alternatives for transcribing noisy clinical documents. Using obstetric ultrasound reports written in regionally inflected medical English common to Indian healthcare settings, we compare eight systems in terms of transcription accuracy, noise sensitivity, numeric accuracy, and computational efficiency. Compact multimodal models consistently outperform both classical and neural OCR pipelines. Despite higher computational costs, their robustness and linguistic adaptability position them as viable candidates for on-premises healthcare digitization.
\end{abstract}

\section{Introduction}
\label{sec:introduction}
Digitization of clinical records increasingly relies on ad-hoc, camera-based document capture rather than controlled scanning \cite{Mosa2012, NETTROUR201957, walters2024smartphone}. In busy healthcare environments, particularly in obstetrics, where large volumes of reports are produced daily, clinicians often photograph printed documents with smartphones to save time and streamline workflows. These images, while convenient, are frequently degraded by blur, uneven illumination, shadows, or physical wear, posing major challenges for text extraction. Robust Optical Character Recognition (OCR) under such noisy, real-world conditions is therefore essential to enable searchable electronic records and downstream analytics.

Beyond immediate clinical use, effective OCR on low-quality images can unlock the vast potential of digitizing \textit{archived printed medical documents}. Many institutions hold years of legacy reports that remain in paper form, limiting their accessibility for research, auditing, or longitudinal analysis. Accurate text extraction from photographed pages enables rapid conversion of these archives into structured, machine-readable data, supporting evidence-based medicine and secondary data use without extensive manual transcription.

Traditional OCR engines such as Tesseract often underperform on handheld captures. In contrast, recent advances in multimodal language models (MLLMs), which couple vision encoders with language decoders, have shown the emerging ability to transcribe text directly from images, potentially bypassing the need for brittle segmentation and preprocessing stages. Yet the reliability of \textit{compact}, locally deployable MLLMs (up to 14B parameters) for document transcription in clinical contexts remains underexplored.

To address this gap, we conduct a systematic evaluation of traditional OCR, neural OCR, and compact multimodal systems on a private corpus of 340 photographed obstetric ultrasound reports. We assess transcription quality using Character Error Rate (CER), Word Error Rate (WER), and numeric accuracy. Our analysis is guided by four research questions:

\begin{itemize}
    \item \textbf{RQ1:} How do compact multimodal language models compare with traditional and neural OCR systems in accurately transcribing noisy clinical images?
    \item \textbf{RQ2:} How does document noise affect transcription accuracy across OCR pipelines and MLLMs, and which no-reference image quality assessment metrics best predict performance degradation?
    \item \textbf{RQ3:} Do multimodal models preserve numeric accuracy when used as OCR engines in clinical data?
    \item \textbf{RQ4:} What are the computational and deployment trade-offs for on-premises, privacy-constrained use?
\end{itemize}

By jointly examining accuracy, noise sensitivity, and computational footprint, this study evaluates whether compact MLLMs can serve as \textit{practical, privacy-preserving OCR alternatives} for healthcare applications, offering actionable guidance for implementing OCR within secure, resource-constrained medical environments.

\section{Related Work}

\paragraph{OCR in Noisy Clinical Settings.}
Classical engines such as Tesseract~\citep{smith2007tesseract} rely on page segmentation and
character models that are highly sensitive to blur, low contrast, and uneven illumination, 
conditions common in handheld captures of printed medical reports ~\citep{ulhasan2016highperformance}. While targeted preprocessing
can help, assumptions of uniform lighting and clean edges often do not hold in practice.

Modern page-level pipelines like PaddleOCR ~\citep{du2020paddleocr} and
docTR~\citep{docTR2021} integrate a learned text detector with a neural recognizer, avoiding explicit binarization and generally improving robustness over classical OCR. These systems still depend on accurate detection and reading-order reconstruction, and performance might degrade with strong blur or compression. Layout-aware stacks such as Surya~\citep{paruchuri2025surya} extend this paradigm with built-in reading order and table extraction, aligning better with end-to-end document parsing needs in
clinical workflows. Advanced end-to-end variants like GOT-GOT-OCR~2.0~\citep{wei2024gotocr} pushes toward unified OCR by integrating transformer-based vision encoding and language decoding in a single model, eliminating the need for modular stages.

General-purpose compact MLLMs (e.g., Qwen-2.5-VL, Phi-4~MM, InternVL)~\citep{bai2023qwen,microsoft2025phi4, intern35} can read text while
reasoning over document layout and content. However, evidence of robustness on noisy, smartphone captures of clinical material is limited ~\citep{trivedi2025benchmarking}; most training/evaluation still targets synthetic or
well-scanned inputs. This motivates our on-premises evaluation of compact MLLMs alongside dedicated OCR pipelines on obstetric report images.

\paragraph{Image Quality and Noise Estimation.}
To quantify readability, no-reference image quality assessment (NR-IQA) metrics can serve as proxies for noise levels, providing a practical means of estimating input degradation that may affect noise-sensitive OCR systems. General-purpose metrics such as BRISQUE~\citep{mittal2012noreference}, NIQE~\citep{mittal2013making}, and PIQE~\citep{venkatanath2015blind} capture perceptual distortion in natural images. Specialized document IQA (DIQA) approaches predict OCR accuracy directly from documents ~\citep{kang2014convolutional,burie2015icdar2015}. More recent work includes DeQA-Doc~\citep{gao2025deqadoc}, which employs multimodal vision-language models to estimate document quality. We examine how well off-the-shelf NR-IQA and DIQA metrics track OCR/MLLM performance in our clinical, smartphone-captured setting, where degradations (blur, shadows, compression) differ from natural-image assumptions.

\section{Methodology}
\subsection{Problem Statement}

The goal of this study is to evaluate whether compact MLLMs (up to 14~B parameters)  
can serve as practical alternatives to both traditional OCR systems and neural pipelines for transcribing noisy clinical document images.  
We formalize the task as \textit{image-to-text transcription}:  
given an input document image $I$, produce a textual output $\hat{T}$  
that closely matches the reference transcription $T$ in terms of character- and word-level edit distance.

\subsection{Data Description}
The full dataset comprises 340 anonymized obstetric ultrasound reports collected from a clinical partner in India. These reports are routinely generated as part of obstetric imaging workflows, where printed summaries of ultrasound examinations are attached to patient charts and then photographed with mobile phones for inclusion in hospital information systems or for clinician-to-patient communication via secure messaging. This pragmatic capture workflow, while efficient, introduces substantial variability in image quality. All reports were originally printed on paper and subsequently photographed under real-world clinical conditions. Common noise factors include (i) blur, (ii) rotation, (iii) uneven illumination or shadow gradients, (iv) reverse-side text bleed-through, and (v) background texture interference, as illustrated in Appendix~\ref{sec:sample}.

To enable detailed quantitative evaluation, we uniformly sampled 60 documents at random from the 340-report corpus for manual transcription and noise annotation, balancing annotation effort with coverage of typical capture conditions. Appendix~\ref{sec:sample} shows that this 60-document subset is comparable to the full corpus in terms of image-level noise indicators, resolution, and file-size distributions (standardized mean differences $|d| < 0.20$; Welch’s unequal-variance $t$-tests, all $p > 0.20$). Appendix~\ref{sec:noise_annot} details the noise-annotation procedure conducted by three trained annotators following a standardized protocol. Krippendorff’s $\alpha$ (ordinal) ranged from 0.62 (blur) to 0.85 (illumination/shadows), indicating moderate to substantial inter-annotator agreement across the five noise indicators.

\paragraph{Linguistic Style.}
In addition to visual noise, the reports exhibit region-specific phrasing typical of Indian medical English, 
such as “cardiac activity is appreciated” or “liquor is adequate”, which differ from 
North American conventions (e.g., “cardiac activity is present”). 
These expressions are semantically equivalent but stylistically distinct, 
and may challenge models whose language priors are trained primarily on Western clinical corpora.

\subsection{Models and Pipelines Used}
\label{sec:models}

Our goal was to compare options that practitioners can realistically deploy in on-premises clinical settings, spanning the major design choices in document OCR: classical OCR, modular neural pipelines with learned detectors, unified end-to-end OCR, and compact multimodal LLMs (MLLMs) that read text directly from images. Selections were guided by (i) widespread use in production or open ecosystems, (ii) public checkpoints with reproducible inference, and (iii) feasibility on a single workstation GPU or CPU. We intentionally focus on \emph{compact} MLLMs (4-14B) rather than frontier models to reflect real latency/VRAM constraints.

We evaluated eight systems across four families:
(i) Classical OCR baseline: Tesseract, which performs page segmentation and line-level recognition with LSTM decoding and no learned detector.
(ii) Modular neural OCR: docTR, PaddleOCR, and Surya.
These systems pair a learned text detector with a neural recognizer\footnote{Surya is a layout-aware neural stack; we restrict it here to page-level text extraction.}
(iii) End-to-end OCR model: GOT-OCR~2.0, which integrates transformer vision encoding and language decoding in a single compact model, targeting diverse page content without modular stages.
(iv) Compact MLLMs: Qwen-2.5-VL-7B, Phi-4~MM (14B), and InternVL3.5-4B, selected to cover a 4B-14B size range and architecture variations.

All systems received identical whole-page RGB images (no binarization, denoising, or cropping). MLLMs were prompted with:
\emph{``You are performing OCR on this document. Transcribe all visible text verbatim as plain text''.} Further details on experimental setup are
provided in Appendix~\ref{sec:setup}.

\subsection{Evaluation Metrics}

Performance was evaluated using standard word- and character-level error rates (WER and CER), computed as normalized edit distances between model outputs and gold transcriptions. 
To capture clinically relevant precision, we further computed a \textit{numeric accuracy rate} ($N_{acc}$), defined as the proportion of numerical tokens in the reference text that are reproduced identically in the model output. 
Let $G = \{g_1, \dots, g_m\}$ denote the set of numeric spans extracted from the gold transcription and $P = \{p_1, \dots, p_n\}$ those extracted from the prediction. 
After aligning $G$ and $P$ using a greedy, order-preserving sequence matcher, numeric accuracy is given by:
\[
N_{acc} = 
\frac{|\{\, (g_i, p_i) \mid g_i = p_i \,\}|}{|G|}.
\]
That is, $N_{acc}$ represents the fraction of numeric spans in the reference text that are reproduced verbatim, serving as a sensitive indicator of clinical reliability. Further details on evaluation protocol are provided in Appendix~\ref{sec:setup}.

\section{Results}

This section presents findings addressing the four research questions introduced in Section~\ref{sec:introduction}. 
All metrics are reported with stratified bootstrap 95\% confidence intervals (10{,}000 iterations) unless otherwise noted. 
For Spearman rank correlations, we report raw $p$-values together with the corresponding FDR-adjusted $q$-values obtained via the Benjamini-Hochberg (BH) procedure.

\subsection{RQ1: Comparative Accuracy of OCR and Multimodal Models}

Table~\ref{tab:cer_wer_main} reports mean WER and CER 
for all systems evaluated on the 60 manually transcribed ultrasound reports.

\begin{table}[h!]
\centering
\resizebox{\columnwidth}{!}{%
\begin{tabular}{lcc}
\toprule
\textbf{Model} & \textbf{CER (95\% CI)} & \textbf{WER (95\% CI)} \\
\midrule
\multicolumn{3}{l}{\textit{Classical OCR}} \\
Tesseract      & 0.189 (0.132, 0.253) & 0.276 (0.217, 0.339) \\
\midrule
\multicolumn{3}{l}{\textit{Neural OCR Pipelines}} \\
PaddleOCR      & 0.111 (0.084, 0.150) & 0.183 (0.155, 0.219) \\
docTR          & 0.108 (0.081, 0.141) & 0.173 (0.146, 0.205) \\
Surya          & 0.135 (0.081, 0.202) & 0.220 (0.160, 0.291) \\
\midrule
\multicolumn{3}{l}{\textit{End-to-End Neural OCR}} \\
GOT-OCR~2.0    & 0.101 (0.074, 0.139) & 0.395 (0.333, 0.463) \\
\midrule
\multicolumn{3}{l}{\textit{Compact Multimodal LLMs}} \\
InternVL-3.5-4B & 0.040 (0.025, 0.064) & 0.096 (0.078, 0.121) \\
Phi-4~MM       & 0.035 (0.018, 0.063) & \textbf{0.075} (0.054, 0.105) \\
Qwen-2.5~VL & \textbf{0.031} (0.023, 0.040) & 0.078 (0.065, 0.093) \\
\bottomrule
\end{tabular}%
}
\caption{Mean Character Error Rate (CER) and Word Error Rate (WER) with 95\% confidence intervals for each system on the evaluation set (lower is better). Best result per column is in bold. Models are grouped by class.}
\label{tab:cer_wer_main}
\end{table}

\begin{figure}[h!]
\centering
\includegraphics[width=1.0\linewidth]{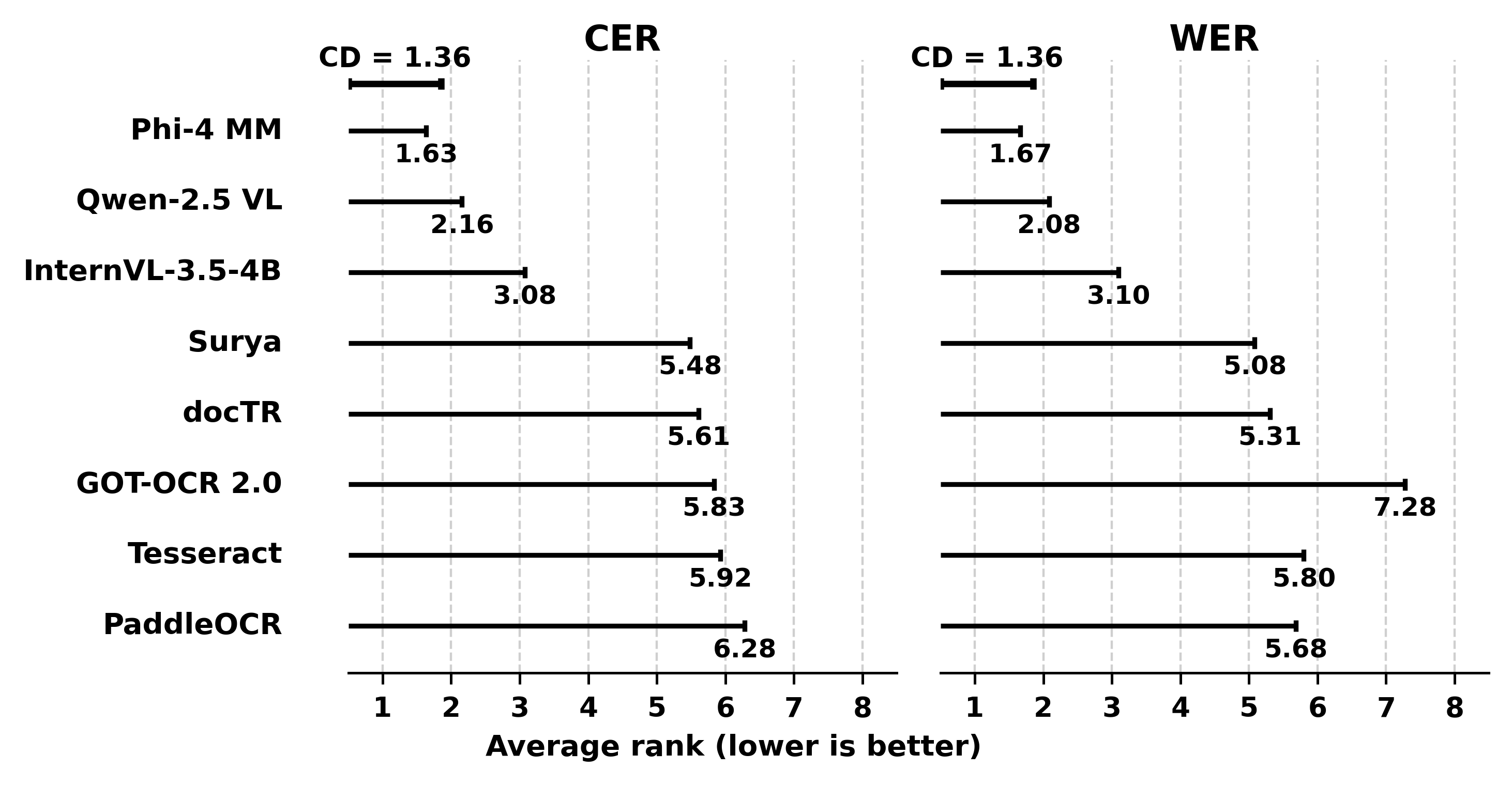}
\caption{Critical-difference diagram of mean ranks computed from 
per-document CER and WER values.  
Lower ranks indicate better performance; groups connected by a bar do not 
differ significantly at $\alpha{=}0.05$.}
\label{fig:cd}
\end{figure}

To assess overall performance differences without assuming a fixed baseline,
we applied the Friedman test to per-document CER and WER values
($N{=}60$, $k{=}8$).
The test revealed a significant effect of model type
for both metrics (CER: $\chi^{2}{F}=251.96$, $p \ll 0.01$;
WER: $\chi^{2}{F}=281.55$, $p \ll 0.01$), confirming that not all systems perform equally. Subsequent pairwise comparisons were conducted using the 
Nemenyi post-hoc procedure, and the resulting mean-rank 
distribution is shown in Figure~\ref{fig:cd}.  

The Nemenyi post-hoc analysis (CD = 1.36 at $\alpha=0.05$) reveals a clear stratification across both CER and WER. The three compact MLLMs form a top-performing group. All remaining systems show statistically indistinguishable performance within a lower tier in terms of CER, confirming that multimodal language models achieve a distinct and consistent advantage over traditional and neural OCR pipelines. Notably, GOT-OCR~2.0 exhibits inflated WER despite relatively low CER. Manual inspection attributes this gap to inconsistent space handling: the model occasionally collapses or inserts spurious spaces, degrading word-level alignment while preserving character-level accuracy.

\subsection{RQ2: Noise Characterization and Model Robustness}
\label{sec:rq2}
\begin{figure}[h!]
\centering
\includegraphics[width=1.05\linewidth]{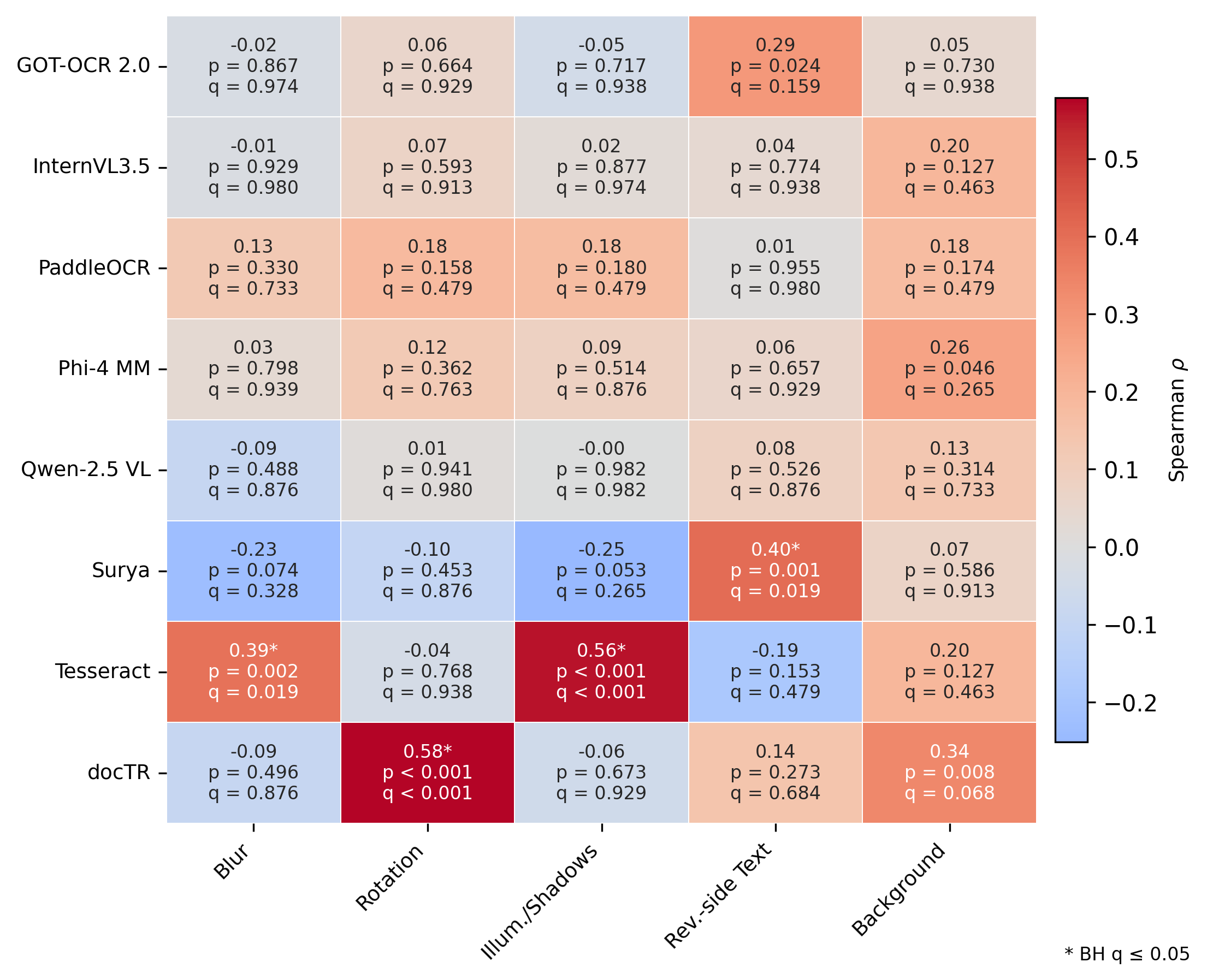}
\caption{Per-model correlations between OCR character error rate and noise indicators after Benjamini-Hochberg correction for multiple comparisons.
Rows correspond to OCR models and columns to noise metrics.
Each cell reports Spearman’s $\rho$ with the corresponding raw $p$-value and FDR-adjusted $q$-value; asterisks mark correlations significant at $q \le 0.05$.
Warmer colors indicate stronger positive associations, while cooler colors denote negative correlations.}
\label{fig:noise_heatmap_models}
\end{figure}

To assess model sensitivity to image noise, we computed per-model Spearman correlations between CER and five manually annotated noise indicators: (i) blur, (ii) rotation, (iii) uneven illumination or shadows, (iv) reverse-side text bleed-through, and (v) background texture interference. The resulting correlation matrix is shown inn Figure~\ref{fig:noise_heatmap_models}.

Noise effects vary substantially across models. Classical and neural OCR pipelines exhibit distinct sensitivities: Tesseract shows strong correlations with blur and illumination or shadow gradients, while docTR is highly sensitive to rotation artifacts. Surya displays significant vulnerability to reverse-side text bleed-through, and GOT-OCR~2.0 shows moderate correlation with this type of noise.
In contrast, compact MLLMs, along with PaddleOCR, demonstrate low and largely insignificant correlations, indicating robustness to the common distortions present in handheld captures.

Manual inspection of the top-five high-CER documents per model supports these patterns: Surya and GOT-OCR~2.0 frequently fail on bleed-through pages, docTR on rotated or skewed layouts, and Tesseract on blurred or shadowed text regions.
Occasionally, MLLMs misalign with the gold transcriptions when background text from another document is visible; human annotators excluded such text from the references, whereas the multimodal models tended to transcribe it, reflecting their broader visual context capture rather than true noise sensitivity. Similar correlation trends for WER are provided in Appendix~\ref{sec:rq2apx}.

\begin{figure}[h!]
\centering
\includegraphics[width=1.05\linewidth]{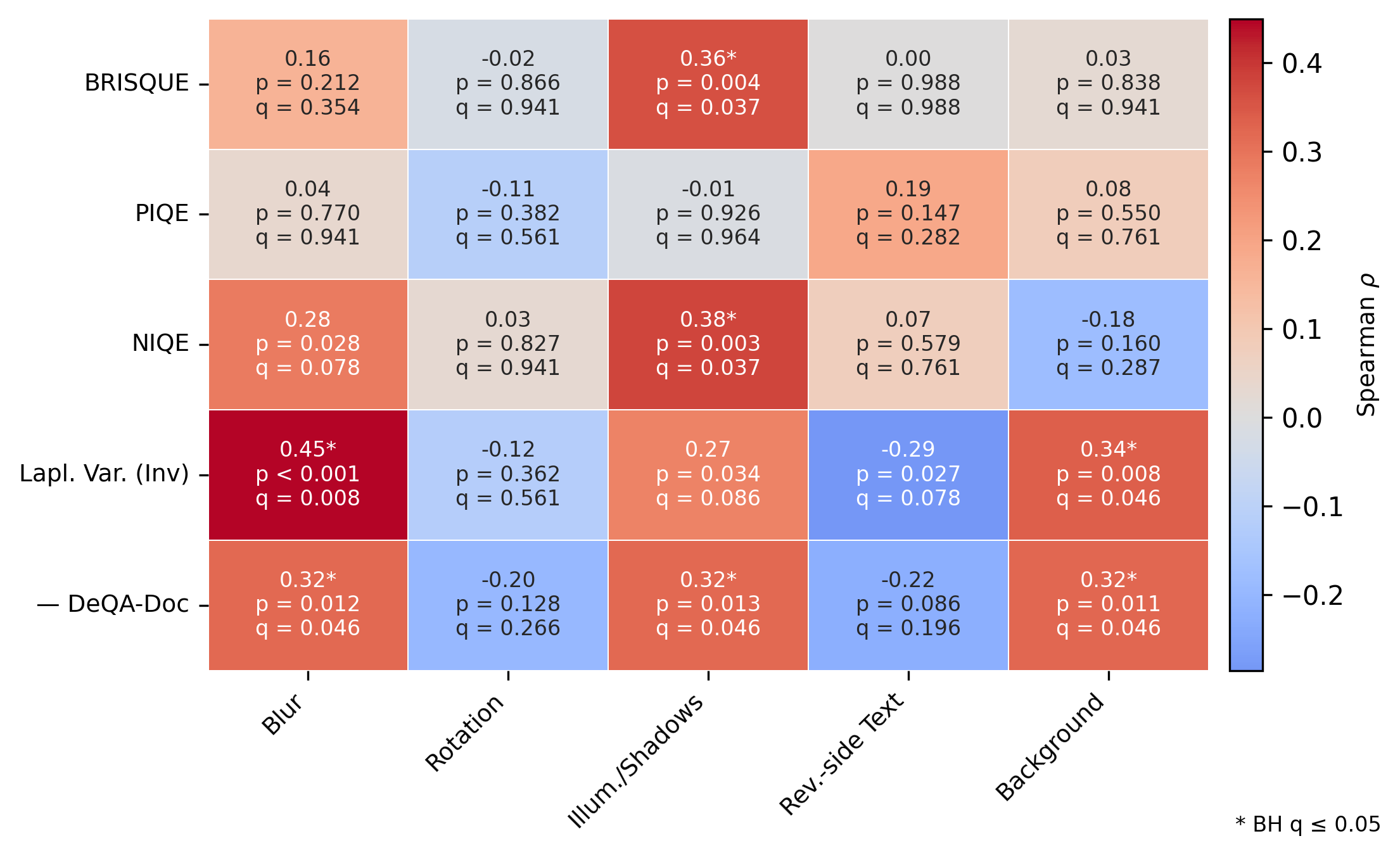}
\caption{Correlations between no-reference image quality assessment (NR-IQA) metrics and manually annotated noise indicators. 
Rows correspond to NR-IQA metrics and columns to noise dimensions. }
\label{fig:noise_heatmap_nriqa}
\end{figure}

\paragraph{NR-IQA Metrics vs. Manual Noise Annotations.}
Figure~\ref{fig:noise_heatmap_nriqa} compares five NR-IQA metrics against the manually annotated noise dimensions. 

Among these, inverse Laplacian variance shows the strongest and most consistent associations, correlating positively with perceived blur and background interference. 
Negated DeQA-Doc\footnote{We negate DeQA-Doc because higher DeQA-Doc scores correspond to better quality of a document.} also aligns well with human ratings, particularly for illumination or shadow gradients, blur, and background noise. NIQE and BRISQUE exhibit significant correlations with illumination or shadow gradients, while PIQE shows no meaningful alignment with the annotated dimensions. For readers interested in correlations between CER/WER and NR-IQA metrics, additional analysis is provided in Appendix~\ref{sec:rq2apx}.

\subsection{RQ3: Numeric Accuracy}

As shown in Table~\ref{tab:numeric_fidelity}, MLLMs achieve over 92\% numeric accuracy, substantially higher than other systems. Appendix~\ref{sec:rq3apx} provides additional details on numeric accuracy with Nemeyni post-hoc analysis confirming that  numeric accuracy is highest and statistically cohesive for the MLLMs. 

To disentangle numeric accuracy from aggregate errors, we examine per-document associations between $N_{\text{acc}}$ and CER/WER, including partial correlations that control for numeric density ($w$, the proportion of characters that are numeric) and document length ($L$, the total number of characters in the document). As summarized in Table~\ref{tab:numacc_correlations}, most systems exhibit strong negative correlations between numeric accuracy and CER/WER, indicating that documents with corrupted numbers also tend to have higher overall error rates. 

\begin{table}[h!]
\centering
\small
\caption{$N_{acc}$ across all models, with 95\% confidence intervals.}
\label{tab:numeric_fidelity}
\begin{tabular}{lcc}
\toprule
\textbf{Model} & \textbf{$N_{acc}$} & \textbf{95\% CI} \\
\midrule
docTR          & 0.884 & [0.842, 0.921] \\
GOT-OCR~2.0    & 0.832 & [0.756, 0.900] \\
PaddleOCR      & 0.674 & [0.620, 0.726] \\
Surya          & 0.821 & [0.778, 0.860] \\
Tesseract      & 0.756 & [0.677, 0.831] \\
InternVL-3.5-4B & 0.927 & [0.889, 0.959] \\
Phi-4~MM       & 0.944 & [0.907, 0.974] \\
\textbf{Qwen-2.5~VL} & \textbf{0.950} & [0.914, 0.979] \\
\bottomrule
\end{tabular}
\end{table}

In contrast, the multimodal language models and docTR show no significant associations after controlling for $w$ and $L$, suggesting that numeric content is largely preserved while residual errors are predominantly non-numeric. Notably, the best-performing model numerically, Qwen-2.5~VL, demonstrates near-zero correlations, confirming its robustness in retaining numerical accuracy independently of overall transcription quality. Additional analysis on correlation between $N_{acc}$ and noise indicators is provided in Appendix~\ref{sec:rq3apx}.

\begin{table}[h!]
\tiny
\centering
\caption{Association between numeric accuracy ($N_{\text{acc}}$) and WER/CER (per-document Spearman $\rho$). Partial correlations control for numeric density $w$ and document length $L$.}
\label{tab:numacc_correlations}
\begin{tabular}{lcccc}
\toprule
 & \multicolumn{2}{c}{\textbf{CER}} & \multicolumn{2}{c}{\textbf{WER}} \\
\cmidrule(lr){2-3}\cmidrule(lr){4-5}
\textbf{Model} & Spearman & Partial\,|\,w,L & Spearman & Partial\,|\,w,L \\
\midrule
\multicolumn{5}{l}{\emph{Classical OCR}}\\
Tesseract      & $-0.594^{*}$ & $-0.674^{*}$ & $-0.646^{*}$ & $-0.688^{*}$ \\
\addlinespace
\multicolumn{5}{l}{\emph{Neural Systems}}\\
docTR          & $-0.188$ & $-0.149$ & $-0.284^{*}$ & $-0.243$ \\
PaddleOCR      & $-0.464^{*}$ & $-0.493^{*}$ & $-0.647^{*}$ & $-0.608^{*}$ \\
Surya          & $-0.458^{*}$ & $-0.444^{*}$ & $-0.442^{*}$ & $-0.447^{*}$ \\
GOT\mbox{-}OCR~2.0 & $-0.580^{*}$ & $-0.696^{*}$ & $-0.521^{*}$ & $-0.666^{*}$ \\
\addlinespace
\multicolumn{5}{l}{\emph{MLLMs}}\\
InternVL\mbox{-}3.5\mbox{-}4B & $-0.239$ & $-0.009$ & $-0.298^{*}$ & $-0.204$ \\
Phi\mbox{-}4~MM & $-0.224$ & $0.025$ & $-0.286^{*}$ & $-0.150$ \\
Qwen\mbox{-}2.5~VL & $-0.043$ & $0.046$ & $-0.113$ & $-0.155$ \\
\bottomrule
\end{tabular}

\vspace{2pt}
\footnotesize $^{*}p{<}0.05$;\; (no star) $p{\ge}0.05$. Partial: Spearman residual correlation after regressing on $w$ and $L$.
\end{table}

\subsection{RQ4: Computational and Deployment Considerations}

Table~\ref{tab:runtime} summarizes latency and memory usage over 60 test images. Appendix~\ref{sec:setup} provides details on hardware and software stack used for experiments.

\begin{table}[h!]
\centering
\caption{Average runtime and memory footprint across 60 test images.
Runtime and memory are reported as mean~$\pm$~SD.
GPU memory denotes peak CUDA allocation; RAM refers to system memory used during preprocessing and inference.}
\label{tab:runtime}
\resizebox{1\linewidth}{!}{
\begin{tabular}{lccc}
\toprule
\textbf{Model} & \textbf{Runtime (s/img)} & \textbf{GPU Mem. (GiB)} & \textbf{RAM (GB)} \\
\midrule
docTR          & 0.81 $\pm$ 0.33  & 1.04  & 5.34  \\
GOT-OCR~2.0    & 4.87 $\pm$ 2.30  & 7.33  & 6.76  \\
PaddleOCR      & 14.14 $\pm$ 3.89 & ---    & 1.00  \\
Phi-4~MM       & 66.79 $\pm$ 38.32 & 47.11  & 7.25  \\
Qwen-2.5~VL    & 54.89 $\pm$ 33.80 & 18.34  & 8.70  \\
InternVL\mbox{-}3.5\mbox{-}4B      & 11.13 $\pm$ 5.04  & 16.75    & 7.10  \\
Surya          & 1.66 $\pm$ 0.80  & 3.84  & 8.51  \\
Tesseract      & 0.63 $\pm$ 0.49  & ---    & 3.01  \\
\bottomrule
\end{tabular}
}
\end{table}

The evaluation underscores tradeoffs in OCR systems for on-premise clinical environments, emphasizing accuracy, efficiency, and resource demands. Compact MLLMs, deliver superior performance but require substantial GPU resources and longer runtimes (11--67~s/img), with Qwen and InternVL needing only around 17--18~GiB (feasible with 20~GB GPUs) while achieving accuracy comparable to Phi-4~MM, making them viable for clinics prioritizing precision despite the hardware needs. Phi-4~MM, in particular, exhibits notable GPU memory variance (not shown in the table), 
consistent with its single-decoder architecture that mixes visual and textual tokens in one context~\cite{microsoft2025phi4}, thus is not recommended for resource-constrined environments. Neural OCR pipelines like docTR (runtime: 0.81~s/img, 1.04~GiB GPU) and PaddleOCR (CPU-only, 14.14~s/img, 1.00~GB RAM) balance moderate accuracy with efficient resource use for general tasks, while Surya (1.66~s/img, 3.84~GiB GPU) offers a similar middle ground. In contrast, the evaluated end-to-end model, GOT-OCR~2.0, showed significantly lower word-level accuracy in this setting, suggesting that some compact transformer-based unified models may remain prone to hallucination noisy conditions. Classical OCR such as Tesseract (0.63~s/img, no GPU, 3.01~GB RAM) remains a strong baseline, often competitive with the evaluated neural models when speed and minimal computational resources are paramount in clinical settings.

\section{Discussion}
\label{sec:discussion}

Compact multimodal LLMs outperformed classical and neural OCR pipelines on 60 noisy obstetric ultrasound reports, achieving the lowest CER and WER while preserving over 92\% numeric accuracy, with no significant partial correlation between numeric accuracy and aggregate errors after controlling for numeric density and document length. In contrast, non-MLLM systems showed numeric accuracy that degraded alongside overall transcription quality, increasing high-risk correction burden in clinical workflows.

Noise sensitivity was pronounced in Tesseract (correlating with blur, shadows, and NR-IQA metrics), but minimal in MLLMs, highlighting their robustness. Qualitatively, MLLMs occasionally transcribed excluded background text, suggesting potential for masks or filters to enhance deployment. Modern NR-IQA metrics are only partially suitable for evaluating document noise and can be a part of low-resource pipelines for document triaging when using classical pipelines like Tesseract that are sensitive to illumination, shadow, and blur, but they cannot consistently capture more specific noise like bleed-through text, rotation, and text in background.

Measured VRAM consumption peaks demonstrate that computational requirements for high-performing MLLMs like Qwen-2.5~VL and InternVL-3.5-4B are accessible with consumer-grade GPUs offering at least 18-20~GB VRAM, unlocking on-premises, privacy-preserving high-quality OCR for clinical environments.

\section*{Conclusion}
Overall, compact MLLMs offer viable privacy-preserving OCR for on-premises clinical use, balancing accuracy and cost. Future work includes structured extraction, layout improvements, and uncertainty-based review loops.

\section*{Ethics Statement}
This study was approved by the institutional Research Ethics Board and conducted in full compliance with institutional and national research ethics guidelines. All obstetric ultrasound reports were de-identified, with patient identifiers removed. Model weights and inference pipelines were deployed entirely on-premises, and no commercial or cloud-based OCR APIs were used. These measures ensured that both data handling and computation adhered to privacy regulations analogous to HIPAA and GDPR. While the original clinical data cannot be publicly shared due to privacy restrictions, a reproducibility package containing the full codebase and experiment results is available at: \url{https://anonymous.4open.science/r/eacl_ind_ocr-C72C/}.

\section*{Limitations}
Our corpus is single-domain (obstetric ultrasound) and region-specific (Indian medical English), which may limit direct portability. We focused on page-level transcription; richer layout preservation and table extraction were not primary endpoints. Finally, compute measurements reflect a single hardware/software stack; absolute latencies may slightly vary across deployments.



\appendix
\section{Dataset Details}
\label{sec:sample}

\subsection{Example Noise Conditions in the Dataset}
\label{sec:noise_examples}

\begin{figure}[h!]
\centering
\begin{subfigure}{0.8\linewidth}
  \includegraphics[width=\linewidth]{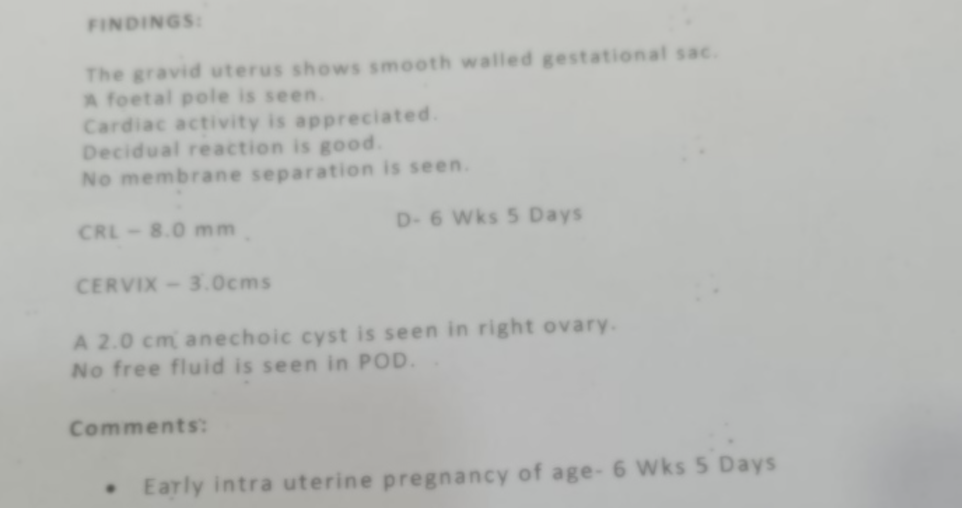}
  \caption{(i) Blur}
\end{subfigure}
\hfill

\begin{subfigure}{0.8\linewidth}
  \includegraphics[width=\linewidth]{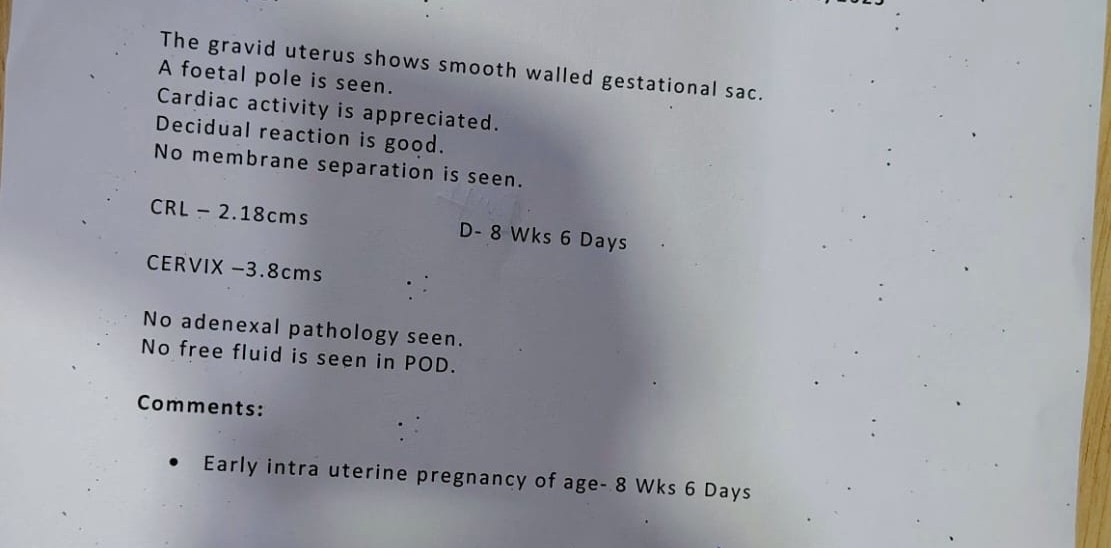}
  \caption{(ii) Rotation}
\end{subfigure}
\hfill

\begin{subfigure}{0.8\linewidth}
  \includegraphics[width=\linewidth]{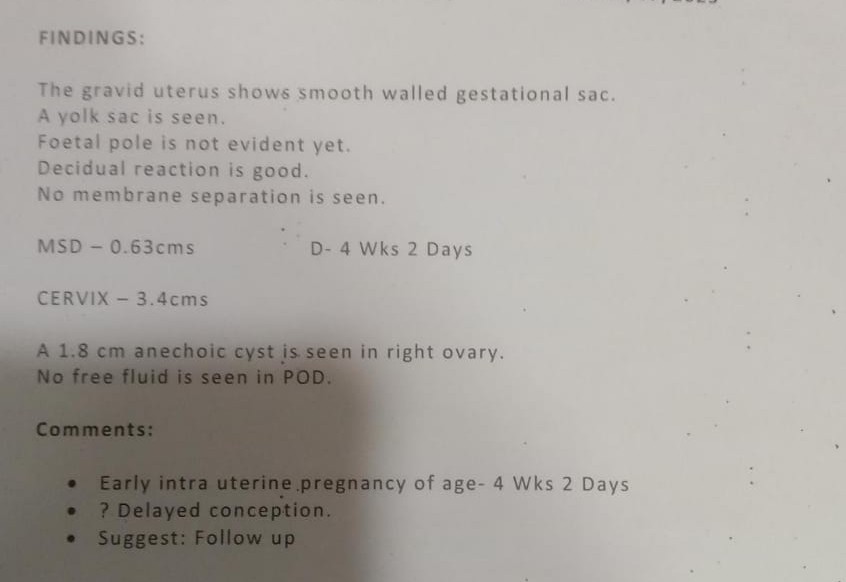}
  \caption{(iii) Uneven illumination or shadow}
\end{subfigure}

\vspace{1em}

\begin{subfigure}{0.8\linewidth}
  \includegraphics[width=\linewidth]{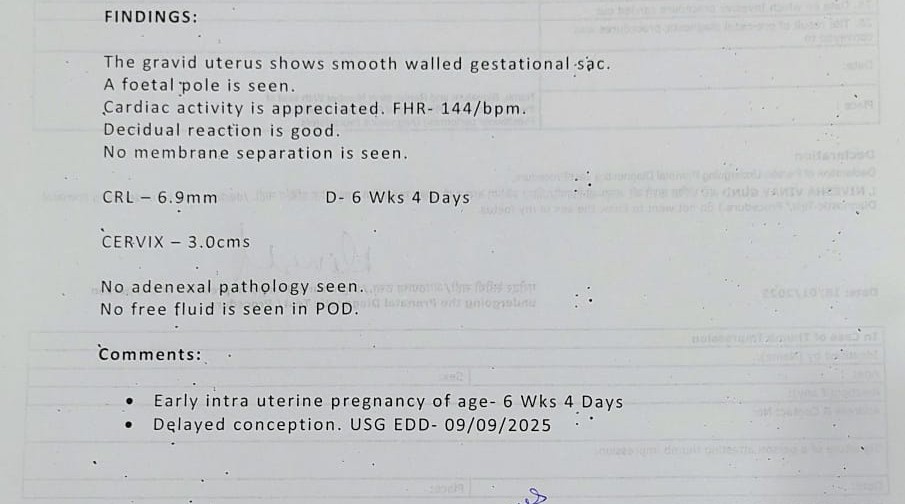}
  \caption{(iv) Reverse-side text bleed-through}
\end{subfigure}

\vspace{1em}

\begin{subfigure}{0.8\linewidth}
  \includegraphics[width=\linewidth]{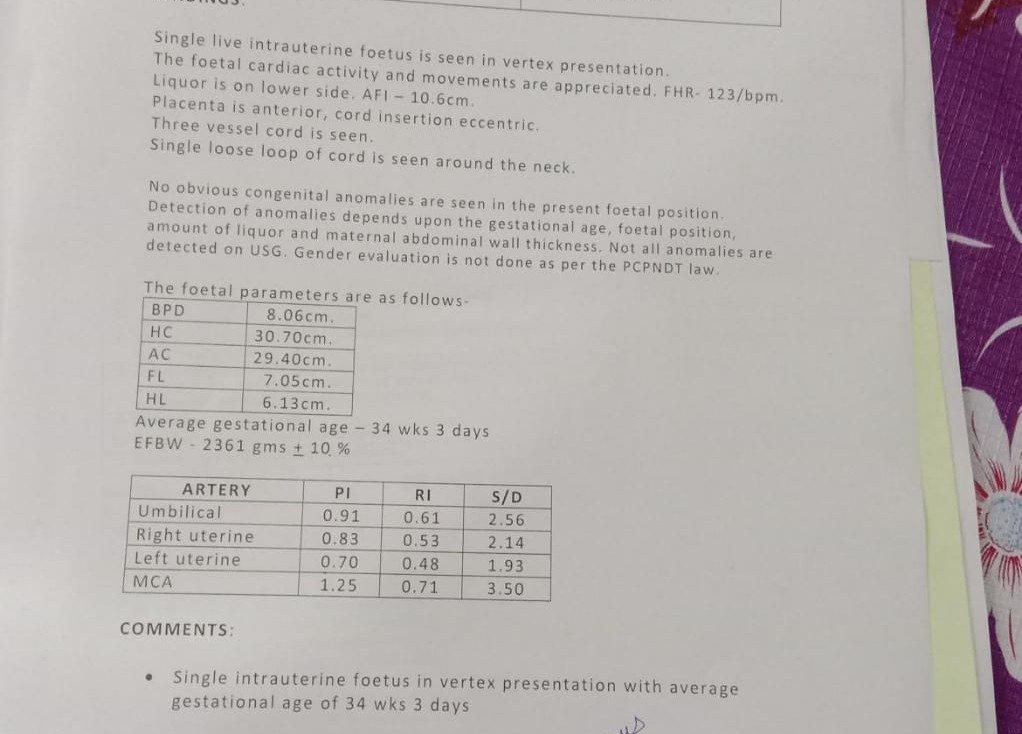}
  \caption{(v) Background texture interference}
\end{subfigure}

\caption{Representative document fragments showing typical noise factors observed in the dataset. These artifacts arise from handheld capture conditions.}
\label{fig:noise_examples}
\end{figure}

Common noise factors in handheld document captures include  
(i) motion blur,  
(ii) uneven illumination or shadow gradients,  
(iii) compression artifacts,  
(iv) reverse-side text bleed-through, and  
(v) background texture interference.  
Representative examples from our dataset are shown in Figure~\ref{fig:noise_examples}.

\subsection{Noise Metric Comparison: Full Dataset vs. Sampled Subset}

Because the 60 manually transcribed reports are a subset of the full corpus, we focus on \textit{practical} representativeness rather than null-difference testing. Table~\ref{tab:noise_stats} summarizes the mean and standard deviation of key noise metrics for both sets. Across all metrics, standardized mean differences were small (Cohen's \(d\), \(|d|<0.20\)). For completeness, Welch's unequal-variance \(t\)-tests found no statistically detectable differences (all \(p>0.20\)). Taken together, the 60-document sample adequately reflects the noise profile of the full corpus.

\begin{table}[h!]
\centering
\caption{Noise metric statistics for the full dataset vs. sampled subset (mean~$\pm$~SD).}
\label{tab:noise_stats}
\resizebox{\linewidth}{!}{
\begin{tabular}{lccc}
\toprule
\textbf{Metric} & \textbf{Full (n=340)} & \textbf{Sample (n=60)} & \textbf{$p$-value} \\
\midrule
BRISQUE & 51.27~$\pm$~28.08 & 47.78~$\pm$~22.03 & 0.247 \\
PIQE & 60.87~$\pm$~8.95 & 59.62~$\pm$~9.64 & 0.353 \\
NIQE & 7.51~$\pm$~1.67 & 7.48~$\pm$~1.77 & 0.905 \\
Laplacian Var. & 8824.97~$\pm$~10106.56 & 7889.08~$\pm$~7972.03 & 0.204 \\
DeQA-Doc & 3.59~$\pm$~0.21 & 3.62~$\pm$~0.21 & 0.400 \\
\bottomrule
\end{tabular}}
\end{table}

\subsection{Image Resolution and File Size Distribution}

Table~\ref{tab:image_stats} reports descriptive statistics for image resolution, file size, and aspect ratio across the full dataset and the evaluated subset. The distributions are closely aligned across all metrics, indicating that the 60-image sample is representative of the overall dataset in terms of basic image characteristics.

\begin{table}[h!]
\centering
\caption{Image size and resolution characteristics of the dataset.}
\label{tab:image_stats}
\resizebox{\linewidth}{!}{
\begin{tabular}{lcc}
\toprule
 & \textbf{Full (n=340)} & \textbf{Sample (n=60)}\\
\midrule
Width (px)  & $977 \pm 200$ & $968 \pm 166$ \\
Height (px) & $1235 \pm 147$ & $1237 \pm 133$ \\
File size (kB) & $97 \pm 77$ & $82 \pm 43$ \\
Aspect ratio (H/W) & $1.32 \pm 0.31$ & $1.32 \pm 0.28$ \\
\bottomrule
\end{tabular}
}
\end{table}

\section{Noise Annotation Details}
\label{sec:noise_annot}

Each document image was rated independently by three annotators along five perceptual noise dimensions:
(i) blur,
(ii) rotation,
(iii) uneven illumination or shadow gradients,
(iv) reverse-side text bleed-through, and
(v) background texture interference.
Annotators assigned an integer score from 1 to 3 for each indicator, using the rubric below.

\begin{itemize}
    \item \textbf{1 (Low / None):} Artifact absent or negligible; the document is easy to read.
    \item \textbf{2 (Moderate):} Artifact present but localized or mild; slight degradation, overall readability preserved.
    \item \textbf{3 (High / Severe):} Artifact clearly visible and substantially affects document readability.
\end{itemize}

Annotators were instructed to rate each noise type independently, ignoring co-occurring distortions, and to rely on visual inspection without pre-processing or enhancement. Prior to the main annotation phase, all annotators jointly reviewed ten representative images covering all five noise types to calibrate their use of the scale.

To quantify inter-annotator reliability on this three-point \emph{ordered} scale, we computed Krippendorff’s $\alpha$ with an ordinal distance function separately for each noise indicator. As shown in Table~\ref{tab:alpha_noise}, values ranged from 0.62 to 0.85, indicating moderate to substantial agreement across annotators, with highest consistency for uneven illumination and reverse-side text bleed-through.

\begin{table}[h!]
\small
\centering
\caption{Inter-annotator agreement per noise indicator, measured using Krippendorff’s $\alpha$ (ordinal).}
\label{tab:alpha_noise}
\begin{tabular}{@{}lc@{}}
\toprule
\multicolumn{1}{c}{\textbf{Noise indicator}} &
\multicolumn{1}{c}{\textbf{$\alpha_\text{ordinal}$}} \\
\midrule
Background texture\\interference & 0.794 \\
Blur & 0.622 \\
Reverse-side text\\bleed-through & 0.851 \\
Rotation & 0.657 \\
Uneven illumination\\or shadow gradients & 0.809 \\
\bottomrule
\end{tabular}
\end{table}

For downstream analyses, we aggregated the three annotator ratings for each document and indicator by taking their arithmetic mean, yielding \emph{one} document-level noise score per indicator in the range $[1,3]$. These aggregated scores are the ones used in all subsequent correlation experiments. Table~\ref{tab:noise_stats} summarizes the distribution of these scores over the 60 annotated documents. On average, images exhibited moderate levels of noise (means between 1.26 and 1.48), with uneven illumination and rotation appearing slightly more frequently than background texture or bleed-through.

\begin{table}[h!]
\small
\centering
\caption{Distribution of aggregated noise ratings across the 60 annotated documents. Scores range from 1 (low/no noise) to 3 (high/severe).}
\label{tab:noise_stats}
\begin{tabular}{@{}lcccc@{}}
\toprule
\textbf{Noise indicator} & \textbf{Mean} & \textbf{SD} & \textbf{Min} & \textbf{Max} \\
\midrule
Background texture\\interference & 1.26 & 0.50 & 1.0 & 3.0 \\
Blur & 1.39 & 0.50 & 1.0 & 3.0 \\
Reverse-side text\\bleed-through & 1.27 & 0.55 & 1.0 & 3.0 \\
Rotation & 1.43 & 0.52 & 1.0 & 3.0 \\
Uneven illumination\\or shadow gradients & 1.48 & 0.56 & 1.0 & 3.0 \\
\bottomrule
\end{tabular}
\end{table}

\section{Experimental Setup Details}
\label{sec:setup}

All experiments were conducted within a secure on‑premises computing environment to ensure that no clinical data or model weights were transmitted outside institutional boundaries. Inference workloads were executed on a workstation equipped with an NVIDIA A100 80GB PCIe GPU (80 GB VRAM), dual‑socket AMD EPYC 7552 processors (96 physical cores, 192 threads), and 1.0 TB system RAM, running Ubuntu 22.04.3 LTS. Table~\ref{tab:software_env} summarizes the core software stack, library dependencies, and model checkpoints used for both OCR and image-quality assessment pipelines. All experiments were implemented in Python~3.10 with PyTorch~2.8 and CUDA~12.8, using mixed-precision inference (\texttt{bfloat16}) where supported. Each model was evaluated via its official checkpoint or inference API to ensure reproducibility and comparability across frameworks.

\begin{table}[t]
\centering
\caption{Software environment and model checkpoints used.}
\label{tab:software_env}
\resizebox{\linewidth}{!}{
\begin{tabular}{lll}
\toprule
\textbf{Scope} & \textbf{Component / Model} & \textbf{Version or ID} \\
\midrule
\multirow{16}{*}{OCR} 
& Python & 3.10.12 \\
& PyTorch & 2.8.0+cu128 (CUDA 12.8) \\
& transformers & 4.57.0 \\
& huggingface\_hub & 0.34.3 \\
& pandas & 2.2.3 \\
& jiwer & 3.1.0 \\
& tqdm & 4.66.5 \\
& Pillow (PIL) & 11.2.1 \\
& pytesseract & 0.3.13 \\
& Tesseract OCR & 4.1.1 (leptonica 1.82.0) \\
& PaddleOCR (Python) & 3.3.0 \\
& GOT-OCR~2.0 checkpoint & \texttt{ucaslcl/GOT-OCR2\_0} \\
& Qwen-2.5~VL checkpoint & \texttt{Qwen/Qwen2.5-VL-7B-Instruct} \\
& Phi-4~MM checkpoint & \texttt{microsoft/Phi-4-multimodal-instruct} \\
& InternVL-3.5-4B checkpoint & \texttt{OpenGVLab/InternVL3\_5-4B} \\
& docTR (python-doctr) & python-doctr 1.0.0 \\
& docTR detector & DB-ResNet50 (pretrained) \\
& docTR recognizer & CRNN-VGG16-BN (pretrained) \\
& Surya OCR & 0.17.0 \\

\midrule
\multirow{10}{*}{Image metrics} 
& Python & 3.10.12 \\
& PyTorch & 2.8.0+cu128 (CUDA 12.8) \\
& torchvision & 0.23.0+cu128 \\
& OpenCV (cv2) & 4.10.0 \\
& pyiqa & 0.1.14.1 \\
& numpy & 1.26.4 \\
& pandas & 2.2.3 \\
& Pillow (PIL) & 11.2.1 \\
& tqdm & 4.66.5 \\
& DeQA-Doc model & \texttt{zhiyuanyou/DeQA-Score-Mix3} \\
\bottomrule
\end{tabular}}
\end{table}

\subsubsection*{OCR Engine Configurations}

\paragraph{PaddleOCR}
\textit{Detector}: \texttt{PP-OCRv5\_server\_det};
\textit{Recognizer}: \texttt{en\_PP-OCRv5\_mobile\_rec};
\textit{Language}: \texttt{en};
\textit{Hardware}: CPU;
\textit{Options}: text-line orientation enabled; default English dictionary; no custom lexicon.

\paragraph{Tesseract}
\textit{Version}: 4.1.1;
\textit{Language}: \texttt{eng};
\textit{Flags}: \texttt{----oem 1} (LSTM engine), \texttt{----psm 6} (single uniform block of text);
\textit{Dictionary}: default;
\textit{User resources}: no user words or patterns.

\paragraph{Evaluation protocol.}

All systems received identical, unprocessed RGB page images loaded with \texttt{PIL}; no binarization or cropping was applied, so each method used its native preprocessing. Model outputs were captured as UTF-8 text.
Before scoring, we normalized both references and hypotheses with the following steps: (1) convert to lowercase; (2) replace newlines and tabs with spaces; (3) remove punctuation (ASCII + common Unicode punctuation); and (4) collapse all whitespace (\verb|\s+|) to a single space and trim.
We then computed CER and WER with \texttt{jiwer}’s character- and word-level metrics on the normalized strings.

Numeric spans were extracted from raw text before any normalization to preserve decimals, signs, slashes, and hyphens using the regular expression \texttt{[+-]?\textbackslash d[\textbackslash d,./-]*} and aligned to compute numeric accuracy rates. 

Each model processed the 60-document evaluation subset, and for each image we recorded wall-clock runtime (\texttt{time.perf\_counter}), process memory (\texttt{psutil} RSS), and GPU memory via NVIDIA NVML (\texttt{pynvml}) when available, falling back to \texttt{torch.cuda.memory\_allocated()}.

\section{Supplementary Noise Correlation Analysis}
\label{sec:rq2apx}

\begin{figure}[h!]
\centering
\includegraphics[width=1.05\linewidth]{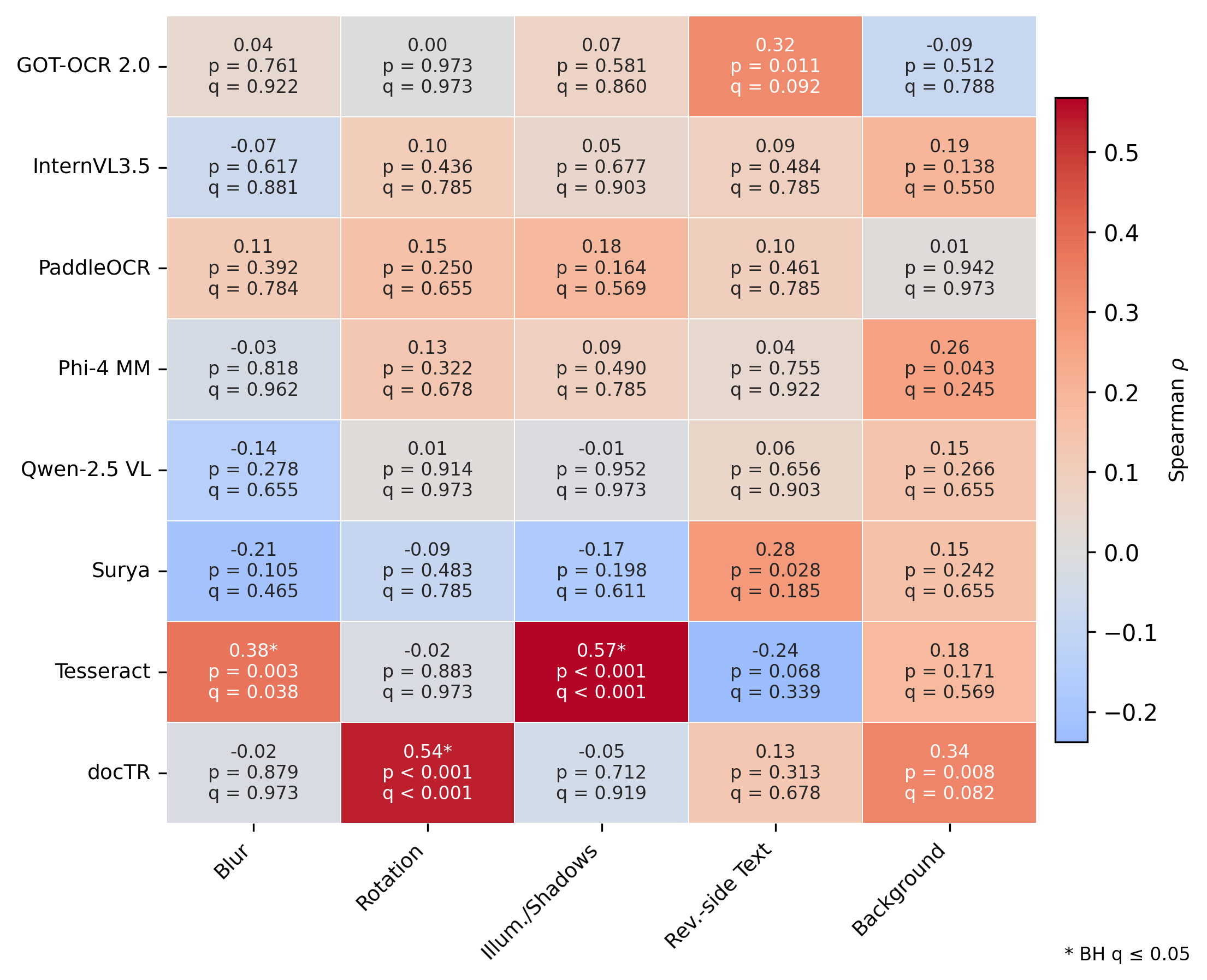}
\caption{Per-model correlations between OCR word error rate and noise indicators after Benjamini-Hochberg correction for multiple comparisons.
Rows correspond to OCR models and columns to noise metrics.
Each cell reports Spearman’s $\rho$ with the corresponding raw $p$-value and FDR-adjusted $q$-value; asterisks mark correlations significant at $q \le 0.05$.
Warmer colors indicate stronger positive associations, while cooler colors denote negative correlations.}
\label{fig:noise_heatmap_models_wer}
\end{figure}

Figure~\ref{fig:noise_heatmap_models_wer} presents per-model correlations between word error rate and the five manually annotated noise indicators. This analysis is provided for reference and complements the CER-based results discussed in Section~\ref{sec:rq2}. Overall, the correlation patterns closely mirror those observed for CER, with classical and neural OCR systems showing higher sensitivity to noise, while multimodal models remain largely unaffected by most noise factors.

\subsection{Supplementary Correlation Between NR-IQA Metrics and OCR Performance}
\label{sec:rq2apx_iqa}

\begin{figure}[h!]
\centering
\includegraphics[width=1.05\linewidth]{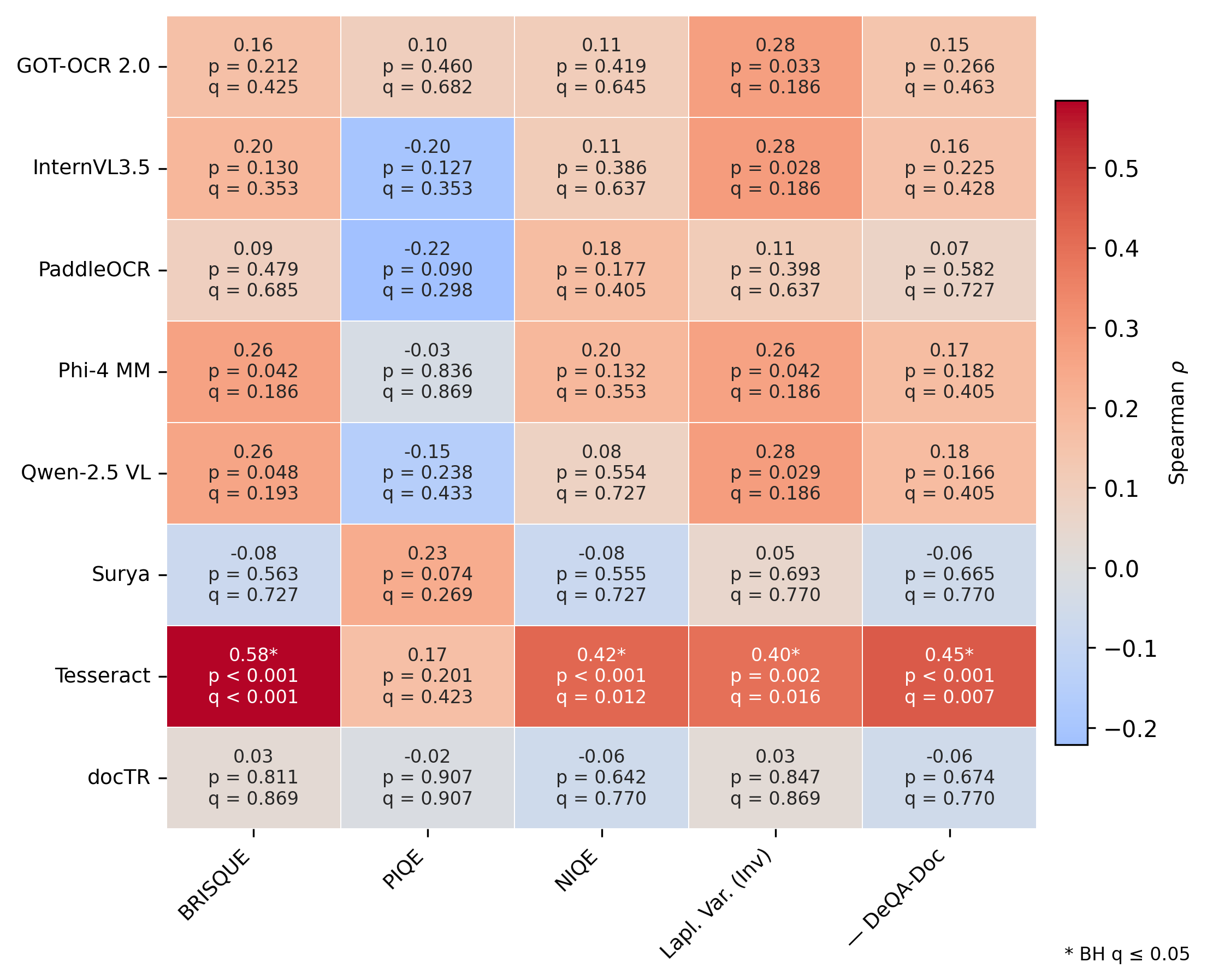}
\caption{Per-model correlations between character error rate (CER) and no-reference image quality assessment (NR-IQA) metrics after Benjamini–Hochberg correction. 
Rows correspond to OCR models and columns to NR-IQA metrics.
Each cell reports Spearman’s $\rho$ with the corresponding raw $p$-value and FDR-adjusted $q$-value; asterisks indicate significance at $q \le 0.05$. 
Warmer colors denote stronger positive correlations.}
\label{fig:heatmap_cer_iqa}
\end{figure}

\begin{figure}[h!]
\centering
\includegraphics[width=1.05\linewidth]{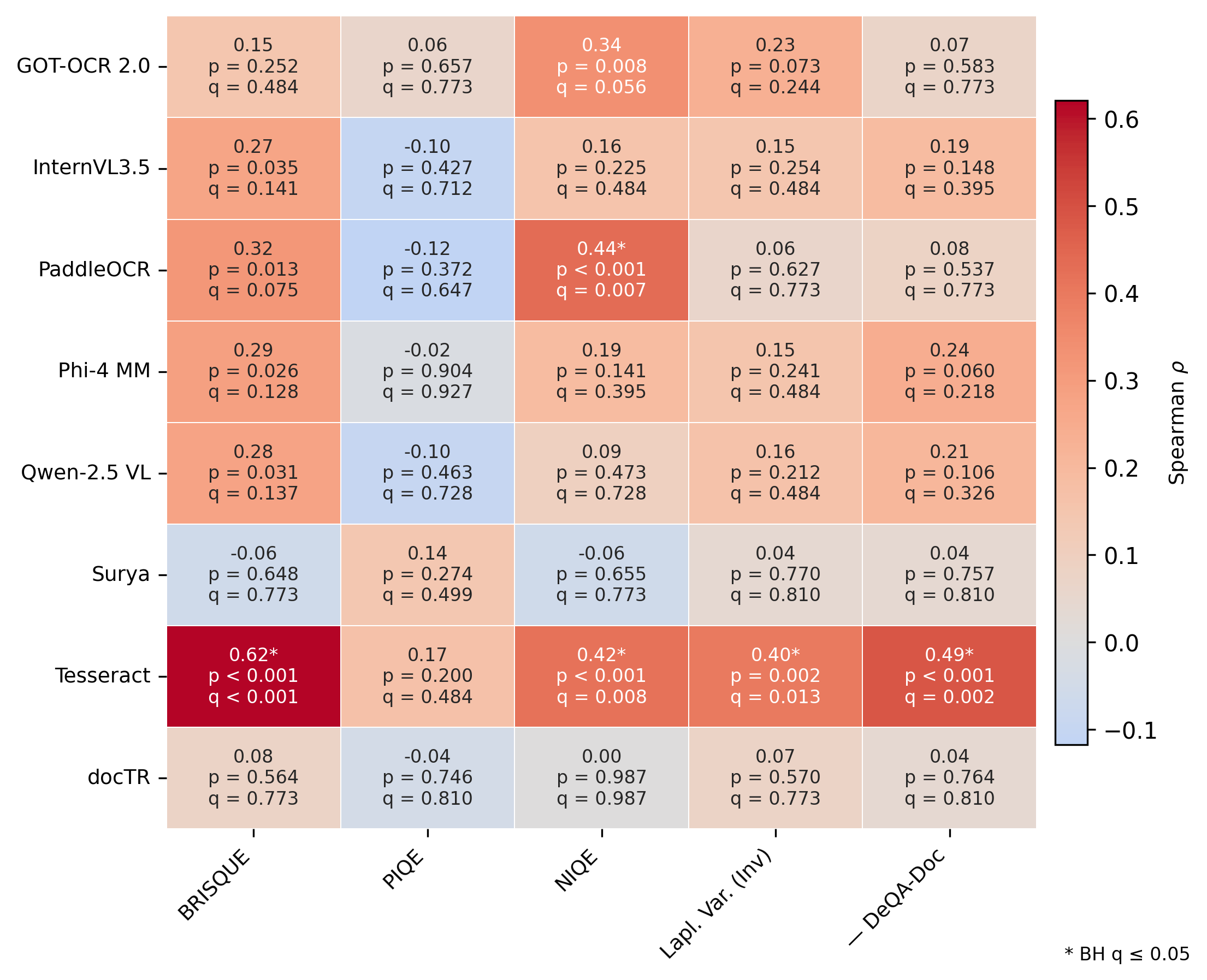}
\caption{Per-model correlations between word error rate (WER) and no-reference image quality assessment (NR-IQA) metrics after Benjamini–Hochberg correction. 
Formatting and interpretation follow Figure~\ref{fig:heatmap_cer_iqa}.}
\label{fig:heatmap_wer_iqa}
\end{figure}

Figures~\ref{fig:heatmap_cer_iqa} and~\ref{fig:heatmap_wer_iqa} summarize correlations between OCR performance (CER and WER) and five NR-IQA metrics across all systems. 

Tesseract exhibits the highest sensitivity to image degradation, showing strong and significant CER correlations with BRISQUE, NIQE, Laplacian variance, and DeQA-Doc ($\rho$ up to 0.62, $q<0.05$), confirming that conventional OCR remains tightly coupled to low-level image quality. PaddleOCR presents moderate WER correlation with NIQE ($\rho=0.44$, $q=0.007$). By contrast, other models show low and nonsignificant correlations across all metrics. 
Overall, these results indicate that NR-IQA metrics are most informative for predicting performance degradation in systems highly sensitive to conventional image noise, such as blur and illumination artifacts. However, they fail to capture more complex, setting-specific degradations including rotation, bleed-through text, and background interference that often characterize real-world clinical documents.

\section{Supplementary Numeric Accuracy Analysis}
\label{sec:rq3apx}

\paragraph{Numeric accuracy across models.}
The critical-difference diagram (CD $=1.36$; $N{=}60$) shows clear stratification in $N_{\text{acc}}$. 

\begin{figure}[h!]
\centering
\includegraphics[width=0.95\linewidth]{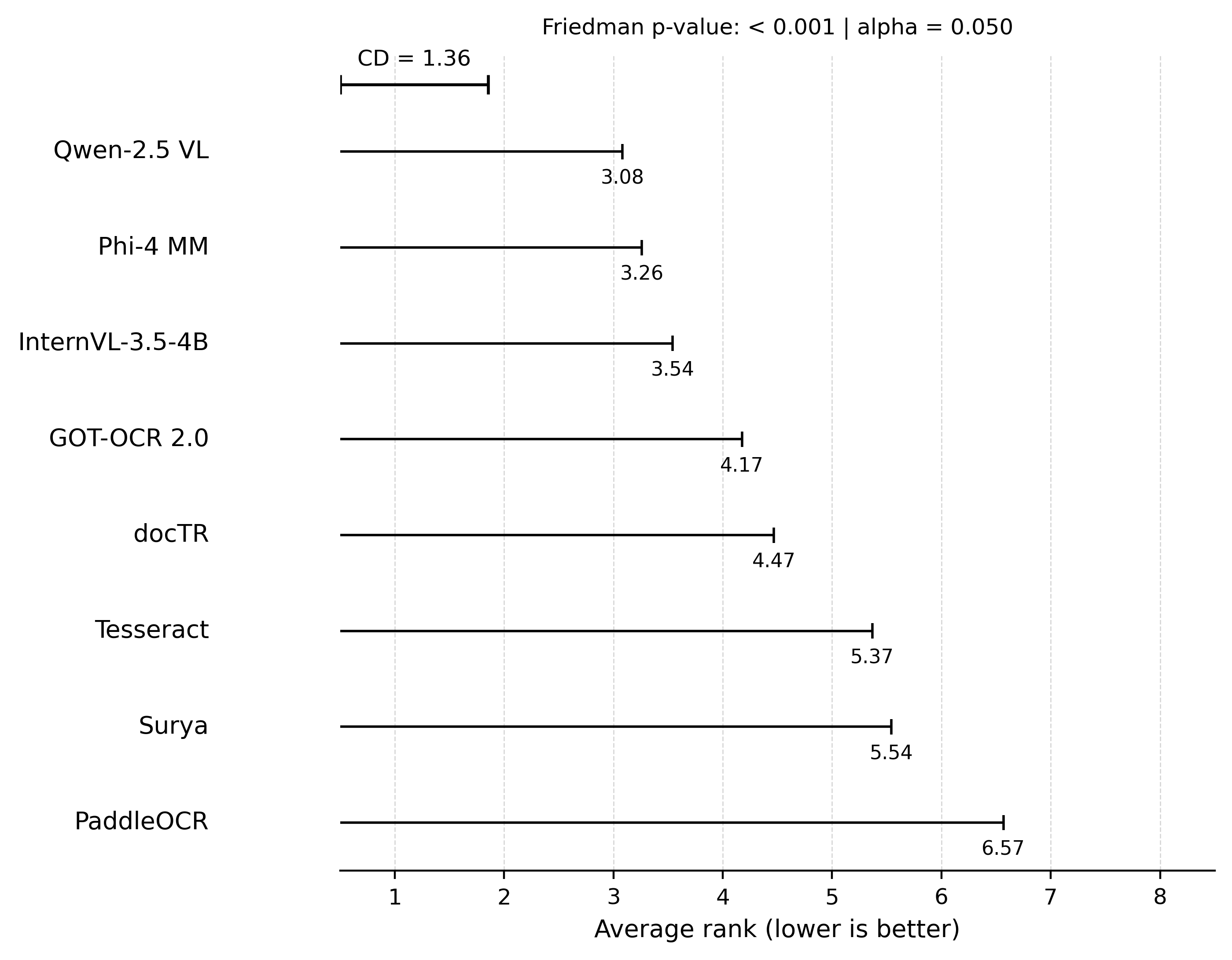}
\caption{Critical difference (CD) diagram for numeric accuracy ($N_{\text{acc}}$) across all models 
($N{=}60$, $\alpha{=}0.05$). 
Lower average ranks indicate better performance. 
Compact MLLMs (Qwen-2.5~VL, Phi-4~MM, InternVL-3.5-4B) form a top-performing group with no significant pairwise differences, 
while GOT-OCR~2.0 and docTR constitute an intermediate tier. 
Tesseract, Surya, and PaddleOCR show significantly lower numeric accuracy.}
\label{fig:cd_numacc}
\end{figure}

The compact MLLMs (Qwen-2.5~VL, Phi-4~MM, InternVL-3.5-4B) form a top group with indistinguishable average ranks. 
GOT-OCR~2.0 and docTR occupy an intermediate band: both are worse than the best MLLM (Qwen-2.5~VL) but not significantly different from Phi-4~MM or InternVL-3.5-4B. 
Tesseract and Surya cluster lower and are significantly worse than the MLLMs; PaddleOCR attains the lowest rank and is significantly worse than the intermediate band (docTR, GOT-OCR~2.0) and all MLLMs, while not distinguishable from Tesseract and Surya. 
Overall, numeric accuracy is highest and statistically cohesive for the MLLMs, with GOT-OCR~2.0 bridging to the neural/classical pipelines below.

\begin{figure}[h!]
\centering
\includegraphics[width=1.0\linewidth]{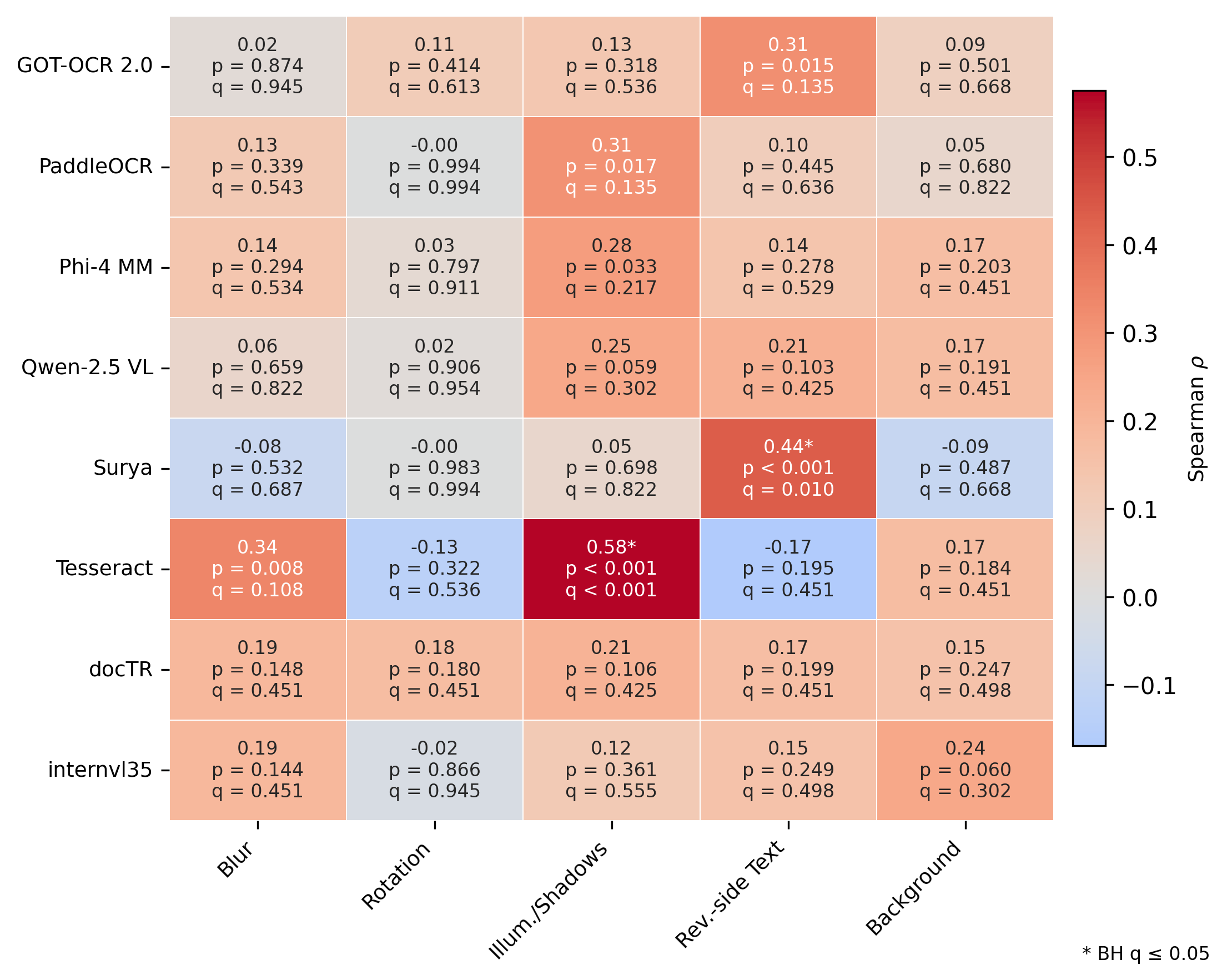}
\caption{Spearman correlations between numeric accuracy ($N_{\text{acc}}$) and
noise indicators across models. Each cell reports Spearman’s $\rho$ with the corresponding raw $p$-value and FDR-adjusted $q$-value. Blue tones denote negative associations
(lower numeric accuracy with increasing noise).}
\label{fig:nemenyi}
\end{figure}

\paragraph{Numeric Accuracy vs. Noise Indicators.}
Figure~\ref{fig:nemenyi} summarizes the correlations between numeric accuracy ($N_{\text{acc}}$) and manually annotated noise dimensions. 
Overall, numeric accuracy remains largely stable across noise types, with few significant associations after Benjamini-Hochberg correction. 
Tesseract shows the strongest sensitivity, with $N_{\text{acc}}$ decreasing under uneven illumination or shadows ($\rho=0.58$, $q<0.001$), consistent with its known fragility to lighting variation. 
Surya exhibits a significant dependence on reverse-side text presence ($\rho=0.44$, $q=0.010$) consistent with overall WER/CER trends for this system. 
All other systems show no BH-significant correlations.

\end{document}